\documentclass[useAMS,usenatbib]{mn2e}
\usepackage{times,mathptmx}
\usepackage{graphicx}
\usepackage{color}
\usepackage{soul}
\sethlcolor{green}

\usepackage{ifpdf}
\usepackage[T1]{fontenc}
\newcommand{\Ion}[2]{#1{\,\sc#2}}
\newcommand{\angstrom}{\mbox{\normalfont\AA}}
\title[Optical transmission photometry of the highly inflated exoplanet WASP-17b]{Optical transmission photometry of the highly inflated exoplanet WASP-17b\thanks{The results presented in this paper are based on observations collected at the European Southern Observatory under programme ID 085.C-0600(A). }}
\author[J. Bento et al.]{\parbox{\textwidth}{J. Bento,$^{1,2}$\thanks{E-mail:
joao.bento@mq.edu.au} P.J. Wheatley,$^{2}$\thanks{E-mail: p.j.wheatley@warwick.ac.uk} C.M. Copperwheat,$^{2,3}$ J.J. Fortney,$^{4}$ V.S. Dhillon,$^{5}$ R. Hickman,$^{2}$ S.P. Littlefair,$^{5}$ T.R. Marsh,$^{2}$ S.G. Parsons,$^{6}$ J. Southworth$^{7}$}\vspace{0.4cm}\\
\parbox{\textwidth}{$^{1}$Department of Physics and Astronomy, Macquarie University, NSW 2109, Australia \\
$^{2}$Department of Physics, University of Warwick, Coventry, CV4 7AL, UK\\
$^{3}$Astrophysics Research Institute, Liverpool John Moores University, Twelve Quays House, Egerton Wharf, Birkenhead, CH41 1LD, UK\\
$^{4}$Department of Astronomy and Astrophysics, University of California, Santa Cruz, CA 95064, USA\\
$^{5}$Department of Physics and Astronomy, University of Sheffield, Sheffield S3 7RH, UK\\
$^{6}$Departmento de F\'{i}sica y Astronom\'{i}a, Universidad de Valpara\'{i}so, Avenida Gran Bretana 1111, Valpara\'{i}so, Chile\\
$^{7}$Astrophysics Group, Keele University, Staffordshire, ST5 5BG, UK\\
}}
\begin{document}

\date{Accepted 2013 October 14.  Received 2013 October 13; in original form 2013 August 20}

\pagerange{\pageref{firstpage}--\pageref{lastpage}} \pubyear{2002}

\maketitle

\label{firstpage}

\begin{abstract}

We present ground-based high-precision observations of the transit of WASP-17b using the multi-band photometer ULTRACAM on ESO's NTT in the context of performing transmission spectrophotometry of this highly inflated exoplanet. Our choice of filters (SDSS $u \rm '$, $g \rm '$ and $r \rm '$ bands) is designed to probe for the presence of opacity sources in the upper atmosphere. We find evidence for a wavelength dependence in the planet radius in the form of enhanced absorption in the SDSS $r \rm '$ band, consistent with a previously detected broad sodium feature. We present a new independent measurement of the planetary radius at $R_{\rm pl} = 1.97 \pm 0.06 R_J$, which confirms this planet as the most inflated exoplanet known to date. Our measurements are most consistent with an atmospheric profile devoid of enhanced TiO opacity, previously predicted to be present for this planet. 

\end{abstract}


\section{Introduction}
\label{Introduction}

\subsection{Atmosphere studies}

The growing sample of transiting planets is proving to be the key towards understanding the structure, composition and formation of exoplanets through analysis of the mass-radius relation \citep[e.g.][]{Pollacco2008}. The extremely low densities of WASP-17b \citep{Anderson2010} and HAT-P-32b \citep{Hartman2011} represent examples of cases that are challenging the current models of planetary formation and bulk structure and it is clear that further observations are needed to better understand these planets. Simple models of core-less planets are not able to reproduce the largest radii for a given mass \citep{Fortney2007, Baraffe2008, Baraffe2010}. Therefore, inflation mechanisms for planetary atmospheres have been proposed that have the potential to explain such low densities \citep[e.g.][]{Burrows2007, Fortney2008,Bodenheimer2000, Jackson2008}, but neither of these mechanisms alone can explain the radii of WASP-17b or HAT-P-32b.

Planetary transits also allow the unique opportunity to sample the absorption profile of the atmosphere of an exoplanet. In wavebands where the opacity of the atmosphere is enhanced due to the presence of a specific absorber, the planetary radius as determined from the transit depth will be larger, and so by making very precise measurements of transit depths we can probe the atmospheric composition and chemistry of the planet (a technique known as \emph{transmission spectroscopy}). The first detection of the wavelength dependence on the radius of an exoplanet was achieved in a narrow band containing the  \Ion{Na}{i} doublet. This observation of the transit of HD\,209458b using the Hubble Space Telescope (HST) \citep{Charbonneau2002} showed an enhanced planet radius in this band, implying a higher opacity of its atmosphere in this wavelength range due to the presence of atomic sodium. The atmosphere of HD\,209458b has also been explored at low-spectral resolution across the optical waveband, also using HST \citep{Vidal2003, Vidal2004, Sing2008a, Lecavelier2008}. The resulting transmission spectrum is dominated by a short wavelength broadband opacity source (around 300-500nm), interpreted as Rayleigh scattering by $\rm H_2$. It is important to note, however, that a higher planet radius at short wavelengths can also be potentially explained by enhanced photochemical absorption in the near UV by $\rm H_2 S$, expected to be abundant in the atmospheres of hot-Jupiters \citep{Zahnle2009}. Moreover, the sharp \Ion{Na}{i} feature appears to be superimposed upon broad Na absorption thought to be a Stark-broadened component of the \Ion{Na}{i} line. The analysis of the \Ion{Na}{I} line absorption as a function of bandwidth by \cite{Vidal2011} yielded a detailed temperature profile for this planet. As a result, the authors find that low temperatures ($< 900 \rm K$) in the lower layers of the atmosphere must lead to sodium condensation, thereby explaining the lack of sodium absorption in the line core with respect to the expected levels for an isothermal atmosphere.

A low-resolution HST transmission spectrum of the other very bright hot Jupiter, HD\,189733b, also shows evidence for a broadband scattering continuum \citep{Pont2008, Sing2011}. However, in this case the scattering particles are larger, perhaps silicate condensates \citep{Lecavelier2008b}. \cite{Pont2013} combine all previous radii measurements of HD189733b and present a complete spectrum of this planet, which shows an almost featureless profile. The authors attribute this fact to the presence of dust and a haze of condensate grains in the atmosphere of this planet, thereby distinguishing it from the profile of the atmosphere of HD209458b. However, high-resolution HST observations around the sodium doublet by \cite{Huitson2012} show that the \Ion{Na}{I} absorption is also indeed present in the atmosphere of this planet, confirming the previous ground-based detection by \cite{Redfield2008}. The authors detect a very narrow sodium doublet line profile, consistent with either high-altitude haze obscuring any broad \Ion{Na}{}, \Ion{K}{} and water features or a low \Ion{Na}{I} abundance hidden underneath the $\rm H_2$ scattering.

Detections from the ground are extremely challenging and, so far, have been achieved only in the narrow bands around the cores of the \Ion{Na}{i} lines \citep[e.g.][]{Snellen2008, Redfield2008, Sing2012}. Other attempts at repeating the results from HST have been consistent with previous observations but dominated by systematic effects \citep[e.g.][]{Narita2005, Knutson2011}. In part, this is because spectrographs are not designed to be photometrically stable, and it is necessary to decorrelate data against a large number of parameters to remove known systematics \citep{Pont2008, Snellen2008, Bean2013}. However, the main features revealed by HST spectroscopy are broadband components and so broadband observations from the ground, via both low-resolution spectroscopy and photometry, can provide further constraints for the composition of these bodies. 

In addition to detections of atmospheric features in absorption, planetary atmospheres have also been detected in emission using secondary eclipse measurements. Observations in the IR using the Spitzer space telescope have been obtained for a number of planets \citep[e.g.][]{Deming2005, Charbonneau2005, Todorov2010, Nymeyer2011, Beerer2011, Deming2011}. These were followed by detections in the optical from space \citep{Snellen2009, Alonso2009, Borucki2009}. Moreover, observations of this phenomenon from the ground have now been performed for several very hot exoplanets, mostly centred in the \emph{K} band \citep[e.g.][]{Sing2009, Lopez2010, Mooji2011, Croll2011} and also in the IR \citep{Mooji2009, Burton2012}. A surprisingly wide range of brightness temperatures have been measured, and this is thought to result from temperature inversions in the atmospheres driving emission at low pressures. 

Early models of exoplanet atmospheres \citep{Hubeny2003, Burrows2007, Fortney2008} predicted that two possible scenarios could take place, depending on incident flux, which would determine whether a temperature inversion in the upper atmosphere is present. Fortney et al. predict that the transmission spectra of highly-irradiated planets should be dominated by TiO opacity, which can be distinguished from Rayleigh scattering because the opacity decreases in the \emph{U} band. Specifically, hot planets (class pM) are predicted to have optical opacities dominated by TiO molecular bands, while in cool planets (class pL) the TiO should have condensed out of the atmospheres. The strong optical TiO opacity results in a temperature inversion in the upper atmosphere of the planet, driving infra-red molecular bands into emission and explaining the high brightness temperatures measured with Spitzer. 
Further research suggests that stellar activity may also be a factor contributing towards the presence of a temperature inversion \citep{Knutson2010}, and that perhaps TiO may be naturally depleted from the upper atmosphere of these planets, since at low enough temperatures it is predicted to condense and thus settle into the lower layers of the atmosphere \citep{Spiegel2009}. Additionally, \cite{Madhusudhan2011} have proposed that there is a connection between the C/O ratio, which is a directly measurable quantity using transmission spectroscopy, and the presence of a temperature inversion. Since most of the oxygen in a carbon rich atmosphere is in the form of CO, the TiO and VO molecule content may be reduced or, in fact, negligible. In any case, further multi-wavelength observations of planetary transits are key towards a more complete picture of this problem.

\subsection{WASP-17b}

The highly inflated transiting planet WASP-17b \citep{Anderson2010} is an example of a planet that falls close to the irradiation transition level between pM and pL class planets. Recent follow-up studies by \cite{Anderson2011} and \cite{Southworth2012} have revealed a planet radius larger than initially measured, confirming this planet as the largest known to date. \cite{Anderson2011} used NASA's \emph{Spitzer} telescope to detect the planet's secondary eclipse at wavelengths of 4.5$\rm \mu m$ and 8$\rm \mu m$. Simultaneously with existing transit light-curves and radial-velocity measurements, they estimate the planet radius to be around 2.0$R_{\rm J}$, raising the question as to what can cause such a high planet radius. Particularly, the low orbital eccentricity observed suggests that tidal heating is unlikely to be responsible for its inflated state. Moreover, \cite{Southworth2012} used further ground-based high-precision defocussed photometry observations to confirm the very high radius of this planet ($R_{\rm pl} = 1.93 \pm 0.05$) but find no evidence of the previously measured orbital eccentricity. Nevertheless, the very low measured density of WASP-17b ($\rho = 0.062 \pm 0.005 \rho_{\rm Jup}$), coupled with a large scale height (see Section \ref{selection}) and high host star brightness, makes it an ideal candidate to perform transmission studies on. 

In this paper we report observations of the transit of this planet using the multi-band fast photometer ULTRACAM \citep{Dhillon2007} on ESO's NTT telescope. In Section \ref{Observations} we describe the observations and strategy employed for this experiment, whilst Section \ref{Data_reduction} outlines the data reduction method used. The data analysis is described in Section \ref{Data_analysis} and we present and discuss the results in Section \ref{Results}. 

\section{Target selection}
\label{selection}

Since wavelength dependent differences are easier to detect in planets with more inflated atmospheres, the atmospheric scale height - the characteristic length scale for a planetary atmosphere - is usually the key parameter for target selection. It is defined as the altitude above which the pressure drops by a factor of $e$, and is derived from first principles based on hydrostatic equilibrium \citep{ExoplanetAtmospheres}. The scale height $h$ as a function of the equilibrium temperature $T$ and surface gravity $g$, where $\mu$ is the mean molecular mass of the particles in the atmosphere and $k$ is Boltzmann's constant is given by

\begin{equation}
\label{h} h = \frac{ k T }{\mu g } .
\end{equation}

Our choice of target was based not only in terms of the large predicted scale height of the atmospheres of these planets, but also taking into consideration the relative size of one atmospheric scale height with respect to the stellar disk, thereby maximising the potentially detectable signal. The numerical expression of the difference in the depth of a planetary transit between multiple wavelengths due to the extra opacity of one pressure scale height $D_{sh}$ is given by

\begin{equation}
\label{scale_ratio} D_{sh} = \frac{\pi (R_{pl} + h)^{2} - \pi R_{pl}^{2}}{\pi R_s^{2}} = \frac{2 R_{pl} h + h^2}{R_s^2} .
\end{equation}

\noindent This parameter is a dimensionless measure of the fraction of the stellar light blocked by the section of the planet's atmosphere responsible for this difference. This equation shows the ratio between the area formed by the scale height annulus and the stellar disk. Assuming that $R_{pl} \gg h$, and using equation \ref{h} for $h$, the result is simply

\begin{equation}
\label{scale_ratio_final} D_{sh} = \frac{2 R_{pl} k T}{\mu g R_s^2} \propto \frac{R_{pl} T}{g R_s^2} .
\end{equation}

\noindent It then becomes evident that the stellar radius has a large impact on the detectability of the signal from a planetary atmosphere and that simply selecting targets based on their scale height is too simplistic. This criterion was used to select targets for observations with ULTRACAM, thereby favouring large scale height planets with deep transits. Other practical restrictions were considered when choosing candidates, such as stellar magnitude and observability. This led to the selection of planets WASP-15b and WASP-17b as the preferred targets. WASP-17b, in particular, has an estimated value of $D_{sh}$ at least twice as large as the two most studied planets in the context of transmission studies, HD209458b and HD189733b. 

We estimate the value of one scale height for WASP-17b to be approximately 2080km (see table \ref{all_params}), which is equivalent to $D_{sh} = 4.7 \times 10^{-4}$ or 3.02\% of the transit depth. A photometric detection of such signal is within the photon limited capability of the instrument we used by a factor of 5 in the optical bands, and previous results (see Section \ref{Introduction}) suggest that the atmospheric features detectable may be equivalent to several scale heights.

\section{Observations}
\label{Observations}

The ULTRACAM instrument \citep{Dhillon2007} consists of a set of 3 frame-transfer CCD cameras that image a field simultaneously through different filters. The incoming beam from the telescope is split using dichroic beamsplitters and directed onto each individual CCD, thereby allowing simultaneous high-precision photometry in 3 separate bands with a negligible read-out time ($\approx 24 \rm ms$). Simultaneous observations of this kind are additionally robust against variability in the system properties, such as stellar activity on time-scales similar to the orbital period of the planet and variability in the characteristics of the planetary system, both in terms of global weather and the presence of moons, which may be different in separate transits.  

\begin{figure}
  \begin{center}
    \includegraphics[width=0.5\textwidth]{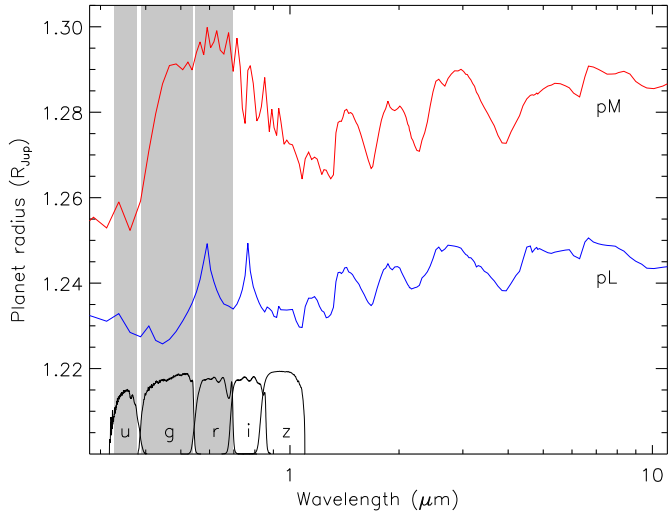}
  \end{center}
  \caption[Exoplanet transmission spectra for two classes with ULTRACAM filters]{\label{Southworth} Predicted planet radius as a function of wavelength for a pL and a pM class planet with a surface gravity of 15\,m\,s$^{-1}$ and a radius of 1.20\,$R_J$ at a pressure of 1\,bar (from \citealt[][Fig. 11]{Fortney2008}). We have indicated the ULTRACAM $u\rm '$, $g\rm '$ and $r\rm '$ passbands with shaded rectangles, as well as the filter response curves. It can be seen that the planet radius varies by over 3\% between $u\rm '$ and $r\rm '$ for the pM planet. The radius variation for the pL planet is dominated by \Ion{Na}{i} absorption (in the $r\rm '$ band).}
\end{figure}

We use a combination of filters that can test the presence of TiO opacity, Rayleigh scattering and \Ion{Na}{i} absorption in the atmosphere of our selected targets. Using the SDSS $u\rm '$, $g\rm '$ and $r\rm '$ filters we can test both the presence of enhanced TiO opacity or a Rayleigh scattering dominated atmosphere from the transit depth differences between the $u\rm '$ and $g\rm '$ filters. This is depicted in Figure \ref{Southworth}, where an example of the atmospheric profiles of a pL and pM class planet are shown along with shaded the regions indicating the filter wavelength bands. Moreover, the $r\rm '$ filter can be used to probe for the presence of a broad sodium absorption feature similar to that detected in the atmosphere of HD209458b \citep{Sing2008a, Sing2008b}. The figure shows atmospheric models for a $R_{\rm pl} = 1.2 R_{\rm J}$ planet with a surface gravity of $g = 15 {\rm m/s}$. However, in the case of WASP-17b ($R_{\rm pl} = 1.95 R_{\rm J}$, $g = 3.16 {\rm m/s}$), the much lower density and larger scale height lead us to expect that any atmospheric features are likely to be more pronounced, and that the variations in radius between the $u \rm '$ and $g \rm '$ bands for a pM class planet can be as high as 8\% of the planet radius.

The observations were performed with the telescope defocused, the amount of which was determined using the model in \cite{Southworth2009b}. We note that the flat-fielding noise component, estimated at 0.1\% per pixel per frame, is significantly reduced once the full-width half maximum (FWHM) of the stellar point spread function (PSF) is larger than 2", and can be essentially negligible if over 100 flat-field twilight frames are used for calibration and accurate guiding takes place. The latter may require continuous monitoring of the PSF position and manual tracking adjustments during the observing run, since flexure between the auto-guider and the instrument causes the PSF to move during the night. Once the defocus amount was determined, the FWHM of the PSF was kept as constant as possible during the night with small adjustments to the telescope focus to compensate for changes in the seeing conditions. The particular observations described in this paper were performed with the telescope defocused to an equivalent FWHM of the PSF of 3-4". This range is large because the focus is not identical between the three arms of ULTRACAM. However, an excessive defocus can also be problematic, as it increases the chances that a star of interest falls on an unwanted detector feature, such as a particularly bad pixel or column. Care was taken to ensure this did not occur. It also increases the chances of stellar blending. We used the SExtractor software \citep{Bertin1996} to estimate the potential contribution of background stars towards the flux from the target. We selected 100 random images taken in the $g\rm '$ band, since this is the arm with the highest sensitivity, and find that, on average, the maximum threshold setting such that a second source is detected by the software is 0.004. This value is an upper limit of the fraction of total light inside the aperture contributed by a second undetected star. 

In this paper we present data of the transit of WASP-17b on the night of 26th April 2010. We used an exposure time of 7.2 seconds over the duration of the night in a 4-window mode arrangement.

\begin{table}
\begin{center}  
\begin{tabular}{ccccc}
\hline
Target & $r \rm '$ & $r \rm '$-$g \rm '$ & RA (h:m:s) & DEC (d:m:s) \\
\hline
WASP-17 & 11.9 & 0.2 & 15:59:50.94 & -28:03:42.34 \\
3UC	124-185767 & 13.03 & 0.03 & 15:59:33.29 & -28:03:23.35\\
3UC	124-185773 	& 14.3 & 0.1 & 15:59:34.54 & -28:03:33.34\\
3UC	124-185780 & 12.8 & 0.2 & 15:59:35.42 & -28:02:38.50\\
3UC	124-185913 & 13.6 & 0.2 & 15:59:53.56 & -28:04:55.48\\
\hline
\end{tabular}
\caption[Magnitude and colour of WASP-15, WASP-17 and comparisons]{\label{star_table} Table containing the $r \rm '$ magnitude, $r \rm '$-$g \rm '$ colours, Right Ascension and Declination (J2000) for the target (WASP-17) and comparison stars used. These are based on information from the NOMAD catalogue \citep{Zacharias2005}. Magnitude conversions done using the method of \cite{Jester2005}.}
\label{colours}
\end{center} 
\end{table}

\section{Data reduction}
\label{Data_reduction}

The calibration and photometry were carried out using ULTRACAM's dedicated software pipeline\footnote{http://deneb.astro.warwick.ac.uk/phsaap/\\software/ultracam/html/index.html} \citep{Dhillon2007}. This is a powerful software suite designed specifically to reduce data from this instrument with a wide variety of options. Biases and flats were used in a standard way.

Each image was analysed and aperture photometry performed on every star of interest. A total of 4 comparison stars was used. Table \ref{star_table} contains the magnitude, colour and coordinates for these. A fixed sized aperture was used throughout the night, large enough to ensure that most of the stellar flux was contained inside the aperture but chosen to minimise the noise from the background contribution. Using the rms of the out-of-transit data, the final used aperture radius was 40 pixels (equivalent to 14"), with the sky annuli between 60 and 85 pixels in radius. Results were found to be insensitive to variations in the aperture size over a minimum threshold (35 pixel radius) where a non-negligible amount of flux from the star would fall outside the aperture at high airmass.The ULTRACAM software pipeline provides the possibility of defining background \emph{masking apertures}. These are regions set manually around each main aperture where the user believes a nearby star is present and ensures that flux from the pixels inside these smaller masks is not considered for background estimation purposes.

\section{Data analysis}
\label{Data_analysis}

The Markov-chain Monte-Carlo (MCMC) fitting routine used is based primarily on the LCURVE fitting procedure written by one of us (TRM), described in greater detail in \cite{Copperwheat2010}. Additionally, these data were fitted using the method of \cite{Copperwheat2013}, which consists of a modified version of LCURVE in which the light-curves of all 3 arms of ULTRACAM are fitted simultaneously, with the option of fixing a selection of parameters between them. This method provides an added level of robustness against systematic effects which may affect data differently between the instrument arms as well as allowing us to achieve precise relative measurements of Rp/a over the three arms. 

Ground-based exoplanet observations using broad-band filters on large telescopes are often plagued by the very small number of available comparison stars to be used. It is often the case that the target is the brightest star visible and a comparison that is a good colour match to the target is almost never available. Therefore, a simple differential photometry procedure with any combination of comparison stars often results in a transit lightcurve with an overall trend that needs to be eliminated, resulting from the wavelength dependent nature of the atmospheric extinction \citep{Smart1933}. This differential extinction effect is often dealt with simply by fitting a second order polynomial to the data outside of transit and interpolating this fit for the entire data set \citep[e.g.][]{Harpsoe2013, Deming2011, Caceres2011, Southworth2012, Maciejewsky2011, Southworth2010b, Siverd2012}. Moreover, spatially dependent systematic effects may have a significant and non-measurable effect on such a fit, which makes the choice of comparison star weighting non-trivial. In the absence of a good colour match of similar brightness to the target within the available set of comparison stars, we have attributed equal weighting to all comparison stars. 

\begin{figure}
    \includegraphics[width=0.5\textwidth]{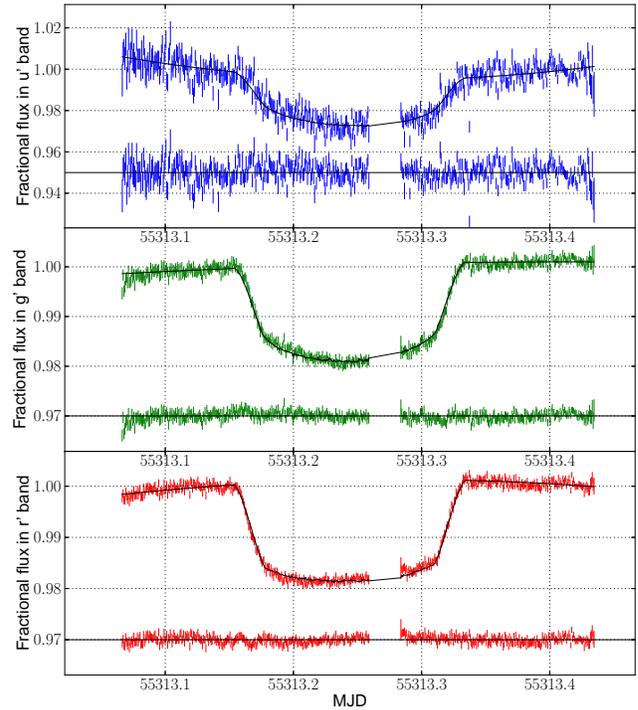}
  \caption{\label{final_fits} Differential magnitude light-curves for the ULTRACAM observations of the transit of WASP-17b taken in the SDSS $u\rm '$ (blue data, top panel), $g\rm '$ (green data, middle panel) and $r\rm '$ (red data, bottom panel). The best fit models and residuals are shown for each case. The data have been averaged into one minute time bins.}
\end{figure}

For the purposes of considering the effects of the polynomial fit to the out-of-transit regions of the light-curve, we elected to include 2 extra parameters (the slope and quadratic terms of the polynomial) in the fitting procedure for the data, which were optimised with the remaining transit parameters. 

We chose to fit the orbital inclination $i$, the time of the transit mid-point $t_0$ and the stellar radius $R_{\ast} /a$ as shared parameters between the 3 light-curves, since these are not wavelength dependent. Additionally, we fit the planet radii scaled to the separation $R_{\rm pl}(\lambda)/a$, the slope and quadratic terms of each polynomial fit ($A(\lambda)$ and $B(\lambda)$ respectively), where $\lambda$ refers to each wavelength band in which observations took place. The model eclipse formulation using the parameters $R_{\ast} /a$ and $R_{\rm pl}/a$ instead of the more commonly used combination of $R_{\rm pl} / R_{\ast}$ and $a$ is due to the fact that LCURVE was designed originally to model binary star eclipses, in which additional parameters related to the accretion disk and bright spots also influence the depth of the several stages of the eclipse. Therefore, a formulation motivated by the timing of the contact points between the various elements of the binary system leads naturally to a combination of parameters scaled to the orbital separation. A similar approach for exoplanet transits is described in \cite{Seager2011}. We use a two parameter Limb-Darkening (LD) law, fitting the coefficient of the 2nd order term of the non-linear law presented as equation 7 in \cite{Claret2000} $u_2(\lambda)$. The 1st order limb-darkening terms $u_1(\lambda)$ are physically motivated based on the available information for the host star and fixed due to their degeneracy with the second order term \citep{Southworth2008}. We used the online data from \cite{Claret2011} through the VisieR catalog to obtain tabulated values for these coefficients, using a star of effective temperature of 1755K, surface gravity of $3.16 \rm m/s^2$ and metallicity of $[Fe/H] = -0.19$. \cite{Copperwheat2013} show that the resulting values of $R_{\rm pl} / a$ are not significantly affected by the choice of limb-darkening law. We chose to fix the mass ratio between the planet and the host star $q$ to the values presented in \cite{Anderson2011} and have imposed a prior constraint on the stellar radius to separation ratio $R_\ast / a$ (degenerate with the planet radius and inclination) in order to obtain a more realistic exploration of the parameter space.

The differential magnitude light curves were initially fitted with a simplex routine to determine an initial model for the MCMC optimisation. The minimisation uses the simplex method via the Numerical Recipes \citep{NumericalRecipes} routine 'amoeba' which works with a group of N+1 points in N dimensional space and moves the points around to move downhill in $\chi^2$. The models were then fed into the MCMC procedure and allowed to iterate until convergence was achieved. Separate parallel chains were created, both with different starting models and perturbed jump-distributions, and were found to yield the same results. Typically, over 300,000 iterations were used for each chain. The final chain contains over 90,000 successful jumps. The fitted light-curves are shown in Figure \ref{final_fits}. Table \ref{results_table} contains the parameter best fit values and uncertainties, estimated as the means and standard deviations of the posterior distributions, together with the assumed values for the first Limb Darkening coefficient. 

We find that the planetary radius is correlated with the orbital inclination and the stellar radius, as seen in \cite{Copperwheat2013}, which motivates the prior constraint on one of the two parameters. However, these two parameters are amongst those chosen to be fixed between all the data from the three arms, and the obtained values are consistent with previous estimates. We are therefore confident that this correlation has not affected significantly our estimate of the planetary radius. We also see a correlation between the planetary radius and the quadratic term $B(\lambda)$ of the polynomial fit to the out-of-transit data in the $u\rm '$ band analysis. Thus, a conservative approach must be taken when interpreting this particular result. It does, however,  highlight the importance of considering the effects of differential refraction when analysing high-precision ground-based photometry of planetary transits.  

The number of degrees of freedom used was 496, equivalent to the number of data points used in the fitting procedure. This is the result of binning all lightcurves into one minute time bins. The total $\chi^2$ of the fit for the three arms respectively are $\chi^2 (u) = 228.16$,  $\chi^2 (g) = 426.58$ and $\chi^2 (r) = 456.31$, which result in reduced $\chi^2_r$ values of $\chi^2_{r} (u) = 0.46$,  $\chi^2_{r} (g)= 0.86$ and $\chi^2_{r} (r)= 0.92$ also respectively. These suggest that the individual error bars for the measurements may be over estimated, particularly in the $u\rm '$ band. However, the effect of systematic errors is likely to the most serious source of uncertainty. In order to gauge the impact of any potential systematic noise in the light curves we have tested the data using a residual permutation analysis method, which consists of iteratively shifting the residuals to the final fits in time and re-fitting the data. The variance of the adjusted model parameters is an indication of the impact of correlated noise. A detailed description of this method is found in \cite{Southworth2008} and we have used the JKTEBOP code developed by one of us (JKT) for this purpose. We find the errors from this analysis to be larger than the those from the MCMC fitting, thus these were adopted.

\section{Results and Discussion}
\label{Results}

\begin{figure}
    \includegraphics[width=0.3\textwidth]{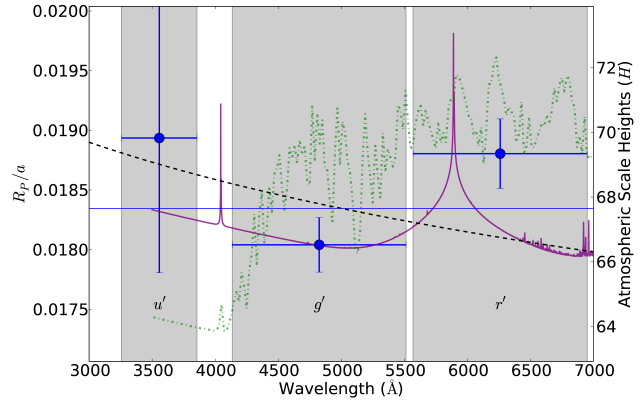}
  \caption[Planet radius as function of wavelength for WASP-17b]{\label{wasp17_radii} Planet radius as a function of observed band (wavelength). The blue filled circles refer to the ULTRACAM data measurements in the 3 bands. The horizontal line indicates the weighted average of the ULTRACAM measurements. We have shaded regions corresponding to the ULTRACAM filters used and show an alternative axis to the planetary radius corresponding to the equivalent number of atmospheric scale heights. We present an example Rayleigh scattering curve as the dashed black line, and models of a pL and pM class planet atmosphere (solid purple and dot-dashed green lines respectively). All 3 models have an offset applied to match the ULTRACAM measurements.}
\end{figure}

The resulting values of this analysis for the radii in all three bands are found to be $R_u / a = 1.89 \pm 0.04 \pm 0.11 \times 10^{-2}$, $R_g / a = 1.80 \pm 0.007 \pm 0.02 \times 10^{-2}$ and $R_r / a= 1.88 \pm 0.006 \pm 0.03 \times 10^{-2}$ for the $u\rm '$, $g\rm '$ and $r\rm '$ bands respectively. The quoted errors for each arm refer to the MCMC and permutation analysis respectively, the latter selected as the final uncertainties. These are plotted in Figure \ref{wasp17_radii}.

It is possible to determine, based on these results, the probability that the radii at the various wavelengths are consistent between them. Since the three measurements shown in Figure \ref{wasp17_radii} in blue have unequal error amplitudes, a statistical test that accounts for this is necessary. The Welch t-test for hypothesis testing was used \citep{Welch1947} for this purpose. This method is a variation of the Student t-test, developed by \cite{Student1908} to assess the statistical significance of the difference between the mean of two distributions, but applied to distributions with unequal variances. For a more in-depth contextualisation in terms of the applications of the Student t-test, see \cite{Venables2002}. We can use the Welch t-test to determine a statistical test quantity $t$ which is an indicator of consistency of each measurement. The statistic is defined as 

\begin{equation}
\label{t} t=\frac{\bar{X}_{1} - \bar{X}_{2}}{\sqrt{\frac{s_{1}^{2}}{N_1} + \frac{s_{2}^{2}}{N_2} }}
\end{equation}

\noindent where $\bar{X}_{i}$, $s_i$ and $N_i$ are mean, variance and sample size of the separate distributions. In order to compare this parameter with tabled probabilities the number of degrees of freedom $\nu$ must also be computed. This is done using the expression 

\begin{equation}
\label{nu} \nu = \frac{ \left( \frac{s_1^2}{N_1} + \frac{s_2^2}{N_2} \right)^2 }{ \left( \frac{s_1^4}{N_1^2 ( N_1 -1 )} + \frac{s_2^4}{N_2^2 ( N_2 -1 )} \right)} .
\end{equation}

\noindent In the context of performing transmission photometry of WASP-17b, we are interested in the difference between the $g\rm '$ band measurement and the others. It is clear from Figure \ref{wasp17_radii} that the $u\rm '$ band measurement is consistent with any of the other values. The relation between the $g\rm '$ and $r\rm '$ bands is not so clear, and the obtained value of t for the comparison between them is $t_{(g,r)}=2.2709$. When compared with tabulated values, we can state that the difference between the values is significant at 90\% confidence, which is typically not enough for a firm detection, but is nevertheless a statistically significant difference. 

The interpretation of these results relies on comparison with existing models of exoplanet atmospheres. Figure \ref{wasp17_radii} also contains examples of such models, where a Rayleigh scattering curve has been plotted, along with models for WASP-17b assuming both the absence of TiO absorption ("pL" class) and presence of such absorption ("pM" class). The Rayleigh scattering model uses the value of 2082 km for the atmospheric scale height, resulting from a combination of the temperature and surface gravities shown in table \ref{all_params}. Additionally, we have assumed Jupiter's mean molecular mass of $\mu = 2.2 {\rm g/mole}$. All three models require an assumption on the planetary radius at 1 bar, which is unknown, and therefore an offset has been applied to match the ULTRACAM measurements. Despite the fact that the three radii are not inconsistent with a featureless transmission spectrum, or indeed with the Rayleigh scattering curve shown, the discrepancy between the $g\rm '$ and $r\rm '$ band radii suggests that most likely scenario is that the higher value for the planet radius in the $r\rm '$ is a result of enhanced opacity, consistent with the broad sodium absorption detected previously in this planet \citep{Wood2011, Zhou2012}. Indeed, a broad sodium absorption feature is predicted to occur for pL class planets, located at $\approx 6000 \rm \angstrom $ (c.f. Figure \ref{wasp17_radii}). If this is the case, the detection of this feature would imply that the atmosphere of WASP-17b is not dominated by dust clouds, which have been suggested to hide opacity sources in the transmission spectrum of exoplanets \citep{Pont2013}. 

\begin{table*} 
\begin{center} 
\caption{Fitted system parameters for WASP-17b. Values forced to be common between all 3 bands are shown only once on the $g\rm '$ band column.} 
\label{results_table} 
\begin{tabular}{lccccl} 
\hline 
Parameter & Symbol & $u\rm '$ & $g\rm '$ & $r\rm '$ \\ 
\hline 
\\
Planet radius (scaled to orbital separation) & $R_{\rm pl} / a $ & $1.89 \pm 0.11 \times 10^{-2}$ & $1.80 \pm 0.02 \times 10^{-2}$ & $1.88 \pm 0.03 \times 10^{-2}$ \\
\vspace{0.3 mm}
Stellar radius (scaled to orbital separation) & $R_{\ast} / a $ & & $0.1409 \pm 0.0003 $ & \\
\vspace{0.3 mm}
Planet/star radius ratio & $R_{\rm pl} / R_{\ast} $ & $0.134 \pm 0.003$ & $0.1277 \pm 0.001$ & $0.1334 \pm 0.002$ \\
\vspace{0.3 mm}
Transit epoch (MJD) (days) & $t_{\rm 0}$ &  & 54558.68535$\pm 0.00008$ &  \\
\vspace{0.3 mm}
Orbital inclination (degs) & $i$ &  & 86.92$\pm 0.04$ &  \\
\vspace{0.3 mm}
Limb Darkening Coefficient 1 (fixed) & $u_1$ & 0.4098 & 0.4393 & 0.2440 & \\
\vspace{0.3 mm}
Limb Darkening Coefficient 2 & $u_2$ & 0.43$\pm 0.07$ & 0.31$\pm 0.02$ & 0.21$\pm 0.02$  \\
\vspace{0.3 mm}
Slope term coefficient & $A$ & $-2.3\pm 0.4 \times 10^{-3}$ & $1.18\pm 0.06 \times 10^{-3}$ & $0.82\pm 0.05 \times 10^{-3}$  \\
\vspace{0.3 mm}
Quadratic term coefficient & $B$ & $8\pm 1 \times 10^{-3}$ & $-0.7\pm 0.2 \times 10^{-3}$ & $-2.0\pm 0.2 \times 10^{-3}$  \\
\vspace{0.3 mm}
\\ 
\hline 
\end{tabular} 
\end{center} 
\end{table*}

The data are unlikely to be consistent with enhanced TiO absorption, which has been suggested to be related to a temperature inversion in the upper atmosphere. In such a case, for a pM class planet the $g\rm '$ and $r\rm '$ band measurements would be expected to be consistent within the quoted errorbars and the planet radius in the $u\rm '$ band should reveal a significantly lower value. This is not observed. We therefore conclude, based on the presented wavelength dependence on the planet radius, that WASP-17b is most likely to have no significant TiO opacity but shows signs of sodium absorption, consistent with previous results. 

The large errorbar of our $u\rm '$ measurement is unable to constrain the atmospheric profile in this part of the optical spectrum. A slightly higher planet radius in the $u\rm '$ band could both indicate a Rayleigh scattering dominated atmosphere or suggest the presence of the previously mentioned enhanced photochemical absorption in the near UV by $\rm H_2 S$, which is predicted to be present in the atmospheres of hot-Jupiters \citep{Zahnle2009}. Further measurements of this kind in the near-UV can determine if this is indeed the case. 

The weighted mean of the three ULTRACAM radius measurements shown in Figure \ref{wasp17_radii} is an independent measurement of this parameter for this planet. It is shown as the blue horizontal line, and is calculated to be $R_{\rm pl} / a = 1.83 \pm 0.04 \times 10^{-2}$. The equivalent value for the planetary radius, not scaled to the orbital separation, is then $R_{\rm pl} = 1.97 \pm 0.06 R_J$. This consistent with the previous measurements by \cite{Anderson2011} and \cite{Southworth2012}.

\begin{table*} 
\begin{center} 
\caption{Summary of system properties. These are a combination of the works of Anderson et al. (2011), for which the reference code is A11, Southworth et al. (2012) (reference code S12) and this paper (reference code B13).} 
\label{all_params} 
\begin{tabular}{lccccl} 
\hline 
Parameter & Symbol & Value & Unit & Reference \\ 
\hline 
\\
Orbital Period & $P$ & $3.7354380 \pm 0.0000068$ & d & A11 \\
Epoch of mid-transit (HJD) & $T_c$ & $2454577.85806 \pm 0.00027$ & d & A11 \\
Transit duration & $T_{14}$ & $0.1830 \pm 0.0017$ & d & A11 \\
Stellar radial reflex velocity & $K_1$ & $53.2 \pm 3.4$ & $\rm m s^{-1}$ & A11 \\
Semimajor axis & $a$ & $0.05125 \pm 0.00099$ & AU & S12 \\
Orbital Inclination & $i$ & $86.92 \pm 0.04$ & $\circ$ & B13 \\
Orbital eccentricity & $e$ & 0.0 (adopted) &  & S12 \\
Planet Mass & $M_{\rm pl}$ & $0.477 \pm 0.033$ & $M_{\rm J}$ & S12 \\
Planet Radius & $R_{\rm pl}$ & $1.97 \pm 0.06$ & $R_{\rm J}$ & B13 \\
Planet equilibrium temperature & $T^{'}_{eq}$ & $1775 \pm 28$ & $\rm K$ & S12 \\
Planet Surface gravity & $g_{\rm pl}$ & $3.16 \pm 0.20$ & $\rm m s^{-2}$ & S12 \\
Planet density & $\rho_{\rm pl}$ & $0.062 \pm 0.005$ & $\rho_{\rm J}$ & S12 \\
Planet/star radius ratio & $R_{\rm pl} / R_{\ast}$ & $0.128 \pm 0.005$ &  & B13,S12\\
Impact parameter & $b$ & $0.37 \pm 0.01$ &  & B13,S12\\
Planet atmospheric scale height & $h$ & $2082 \pm 131$ & $\rm km$ & B13 \\
Star Mass & $M_{\ast}$ & $1.286 \pm 0.076$ & $\mathrm{M_{\odot}}$ & S12 \\ 
Star Radius & $R_{\ast}$ & $1.553 \pm 0.030$ & $\mathrm{R_{\odot}}$ & B13,S12 \\ 
Star surface gravity & $\log g_{\ast}$ & $4.149 \pm 0.014$ & (cgs) & S12 \\ 
Star effective temperature & $T_{eff}$ & $6650 \pm 80$ & $\rm K$ & A11 \\ 
Star density & $\rho_{\ast}$ & $0.324 \pm 0.012$ & $\mathrm{\rho_{\odot}}$ & S12 \\ 
Age &  & $2.7^{+0.6}_{-1.0}$ & Gyr & S12 \\ 
\\ 
\hline 
\end{tabular} 
\end{center} 
\end{table*}

Models of hot-Jupiter atmospheres have struggled to explain very inflated planetary radii, thought to be due to enhanced planetary temperature. Several mechanisms have been put forward (see Section \ref{Introduction}). However, recent studies indicate that, although these can explain the vast majority of known radii of close-in hot-Jupiters, they alone are unable to account for the anomalous radius of planets such as WASP-17b or TrES-4b, partly because of the low mass but also due to a very low eccentricity, not conducive with tidal heating. Instead, other more elaborate scenarios have been suggested, where a combination of tidal dissipation and surface winds induced by stellar irradiation are required to reach the huge sizes of these planets \citep{Leconte2010}. Alternatively, \cite{Laughlin2001} suggest that magneto-hydrodynamic coupling between the planetary magnetic field and the surface flow of particles coupled with ohmic dissipation could cause a larger radius. 

While it is important to note that all previous measurements of the radius of WASP-17b agree with the simpler models of high irradiation and moderate tidal dissipation within the 95\% confidence region of quoted variance, a $\approx 2 R_J$ planet needs to be explained, since our improved measurement in the $r\rm '$ band confirms WASP-17b as the largest known planet and one of the most interesting candidates to perform atmospheric studies on.

\section{Summary}
\label{summary}

We have presented high-precision photometric observations of the transit of WASP-17b in the SDSS $u\rm '$, $g\rm '$ and $r\rm '$ bands using the ULTRACAM instrument on the NTT. We have included 2 extra parameters in our MCMC fit to account for differential extinction variations between the target and comparison stars. We have also fitted the photometry from all 3 bands simultaneously, with the additional constraint that the mid-transit time and orbital inclination should be common between them. 

We find evidence for wavelength dependence of the planet radius, with the corresponding radii estimated to be $R_u / a = 1.89 \pm 0.05 \times 10^{-2}$, $R_g / a = 1.80 \pm 0.02 \times 10^{-2}$ and $R_r / a= 1.88 \pm 0.03 \times 10^{-2}$ for the $u\rm '$, $g\rm '$ and $r\rm '$ bands respectively. The weighted average of the values leads to an improved measurement of the average radius of WASP-17b, which when combined with the estimate of the orbital separation $a$ results in $R_{pl} = 1.97 \pm 0.06 R_J$, confirming this planet as the lowest density exoplanet known to date. Table \ref{all_params} shows an up-to-date summary of relevant parameters for this planetary system.

The 3 ULTRACAM measurements are consistent within the errorbars with a featureless transmission spectrum, but the distribution of the planet radius measurements suggests that WASP-17b is most likely closer to a "pL" class planet in that an enhanced opacity due to TiO is not observed, which was predicted to take place in the atmosphere of this planet. The higher planet radius in the $r\rm '$ band is also consistent with the previously detected broad sodium feature on this planet. Further observations of this system are required to confirm this scenario.

\section*{Acknowledgements}

The authors thank the anonymous reviewer for useful comments that have greatly improved the quality of the manuscript. JB is currently funded by the Australian Research Council Discovery Project DP120103751. CMC, PJW and TRM acknowledge the support of the grant ST/F002599/1 from the Science and Technology Facilities Council (STFC). ULTRACAM, VSD and SPL are supported by STFC grants PP/D002370/1 and PP/E001777/1. JS acknowledges financial support from STFC in the form of an Advanced Fellowship. SGP acknowledges support from the Joint Committee ESO-Government of Chile. The target selection process made use of the SIMBAD data base, operated at CDS, Strasbourg, France.

\input{references.cls}

\bibliographystyle{mn_new}
\bibliography{bibtex}

\bsp

\label{lastpage}

\end{document}